%% file: main.tex
\pdfoutput=1
\documentclass[sigconf]{acmart}
\usepackage{multirow}
\usepackage{listings}
\usepackage{amsmath}
\usepackage{makecell}

\usepackage{enumitem} 
\usepackage{pifont}
\usepackage[dvipsnames, table]{xcolor}
\usepackage{framed}  
\usepackage{hyperref}
\usepackage{stfloats}

\AtBeginDocument{%
  }

\copyrightyear{2026}
\acmYear{2026}
\setcopyright{cc}
\setcctype{by}
\acmConference[KDD '26]{Proceedings of the 32nd ACM SIGKDD Conference on Knowledge Discovery and Data Mining V.2}{August 09--13, 2026}{Jeju Island, Republic of Korea}
\acmBooktitle{Proceedings of the 32nd ACM SIGKDD Conference on Knowledge Discovery and Data Mining V.2 (KDD '26), August 09--13, 2026, Jeju Island, Republic of Korea}
\acmDOI{10.1145/3770855.3818480}
\acmISBN{979-8-4007-2259-2/2026/08}

\theoremstyle{definition} 
\newtheorem{definition}{Definition}[section]

\definecolor{lightgray}{gray}{0.92}
\colorlet{shadecolor}{lightgray}

\newenvironment{promptbox}[1]{
  \smallskip
  \begin{shaded}
  \noindent\textbf{#1}\\
  \rule{\linewidth}{0.4pt}\smallskip 
}{
  \end{shaded}
  \smallskip
}

\lstdefinestyle{jsonstyle}{
    basicstyle=\footnotesize\ttfamily\color{black},
    breaklines=true,
    frame=none,               
    backgroundcolor={},       
    xleftmargin=5pt,
    postbreak=,               
}

\begin{document}

\title{SHERLOCK: Towards Dynamic Knowledge Adaptation in LLM-enhanced E-commerce Risk Management}

\author{Nan Lu}
\authornote{Both authors are corresponding authors.}
\orcid{0009-0007-0731-4343}
\email{lunan5@jd.com}
\affiliation{
  \institution{Beijing Jiaotong University}
  \city{Beijing}
  \country{China}
}
\affiliation{
  \institution{JD.com}
  \city{Beijing}
  \country{China}
}

\author{Yurong Hu}
\orcid{0009-0008-8997-0543}
\email{huyurong7@jd.com}
\affiliation{
  \institution{JD.com}
  \city{Beijing}
  \country{China}
}

\author{Jiaquan Fang}
\orcid{0009-0003-8133-831X}
\email{fangjiaquan1@jd.com}
\affiliation{
  \institution{JD.com}
  \city{Beijing}
  \country{China}
}

\author{Yan Liu}
\orcid{0009-0005-6564-7678}
\email{liuyan961@jd.com}
\affiliation{
  \institution{JD.com}
  \city{Beijing}
  \country{China}
}

\author{Rui Dong}
\orcid{0009-0009-1685-3028}
\email{dongrui_0427@seu.edu.cn}
\affiliation{
  \institution{Southeast University}
  \city{Nanjing}
  \state{Jiangsu}
  \country{China}
}
\affiliation{
  \institution{JD.com}
  \city{Beijing}
  \country{China}
}

\author{Yiming Wang}
\orcid{0009-0009-5144-7930}
\email{22321335@zju.edu.cn}
\affiliation{
  \institution{Zhejiang University}
  \city{Hangzhou}
  \state{Zhejiang}
  \country{China}
}
\affiliation{
  \institution{JD.com}
  \city{Beijing}
  \country{China}
}

\author{Rui Lin}
\orcid{0009-0000-6781-455X}
\email{linrui@jd.com}
\affiliation{
  \institution{JD.com}
  \city{Beijing}
  \country{China}
}

\author{Shaoyi Xu}
\authornotemark[1]
\orcid{0000-0001-5737-2874}
\email{shyxu@bjtu.edu.cn}
\affiliation{
  \institution{Beijing Jiaotong University}
  \city{Beijing}
  \country{China}
}

\renewcommand{\shortauthors}{Nan Lu et al.}

\begin{abstract}
Effective e-commerce risk management requires in-depth case investigations to identify emerging fraud patterns in highly adversarial environments. However, manual investigation typically requires analyzing the associations and couplings among multi-source heterogeneous data, a labor-intensive process that limits efficiency. While Large Language Models (LLMs) show promise in automating these analyses, their deployment is hindered by the complexity of risk scenarios and the sparsity of long-tail domain knowledge. To address these challenges, we propose \textsc{Sherlock}, a framework that integrates structured domain knowledge with LLM-based reasoning through three core modules. 
First, we construct a domain Knowledge Base (KB) by distilling structured expertise from heterogeneous knowledge sources. 
Second, we design a two-stage retrieval-augmented generation strategy tailored for case investigation, which combines input contextual augmentation with a Reflect \& Refine module to fully leverage the KB for improved analysis quality. 
Finally, we develop an integrated platform for operations and annotation to drive a self-evolving data flywheel. By combining real-time hotfixes through KB updates with periodic logic alignment via post-training, we facilitate continuous system evolution to counteract adversarial drifts. 
Online A/B tests at JD.com demonstrate that \textsc{Sherlock} achieves an 82\% Expert Acceptance Rate (EAR) and a 386.7\% increase in daily investigation throughput. 
An additional 90-day evaluation shows that the flywheel successfully recovers from performance decay caused by 
changing tactics twice, raising the EAR ceiling by around 3.5\% through autonomous model updates.
\end{abstract}



\maketitle
\input{sections/introduction}
\input{sections/related_work}
\input{sections/problem_formulation}
\input{sections/methods}

\input{sections/experiments}
\input{sections/conclusion}

\clearpage
\bibliographystyle{ACM-Reference-Format}
\bibliography{main}

\appendix
\input{sections/appendix}
\end{document}

%% file: sections/introduction.tex
\section{Introduction}

The rapid expansion of the e-commerce industry has intensified the adversarial dynamics between shadow economy actors and risk management teams. At JD.com, one of the world's largest e-commerce platforms, risk analysts are tasked with conducting deep investigations into suspicious cases to identify emerging fraud patterns. This process is essential for enhancing both preemptive risk management and post-hoc governance. However, the sheer volume of case analyses imposes a substantial workload on analysts, as each case requires the integration of accumulated expert experience and meticulous scrutiny across multi-source heterogeneous data.

\begin{table*}[t]
  \caption{Case Studies of the Discrepancy between Zero-Shot LLM Reasoning and Domain Expert Rationale.}
  \label{tab:failure_modes}
  \small
  \centering
  \renewcommand{\arraystretch}{1.25} 
  \begin{tabular}{p{9cm}|p{8cm}}
    \toprule
    \rowcolor[gray]{0.95} 
    \textbf{Zero-shot LLM Reasoning} & \textbf{Domain Expert Rationale} \\ \midrule
    \textbf{Observation}: User purchased 9 pairs of pants (XL, XXL, S) in 3 days. \newline
    \textbf{Logic:} Typical households maintain consistent sizing; multi-size bulk buying deviates from personal use. \ding{51} & 
    Significant size inconsistency within a narrow window is a robust indicator of non-personal consumption. \\ \hline
    
    \textbf{Observation:} Multiple purchases of ``Treasure Island'' items detected. \newline 
    \textbf{Logic:} ``Treasure Island'' is associated with high-value collectibles or rare goods. The bulk acquisition of such items typically characterizes speculative behavior and market manipulation. \ding{55} & 
    ``Treasure Island'' is a specialized auction platform of JD.com. It provides auction services for second-hand goods, clearance items, surplus inventory, and special products. Since the unit value of these items is low, bulk purchasing is typically a reasonable behavior. \\ \hline
    
    \textbf{Observation:} 20 delivery orders from the same restaurant sharing identical IPs. \newline 
    \textbf{Logic:} Identical IP addresses for multiple accounts are a definitive indicator of automated bot activities or coordinated fraudulent behavior. \ding{55} & 
    In Local Life Service scenarios, orders predominantly originate from the local service area of physical stores; thus, IP clustering is a common benign phenomenon, particularly for collective orders from campuses or office buildings. \\ \hline
    
    \textbf{Observation:} 80 orders delivered across multiple provinces. \newline 
    \textbf{Logic:} High-frequency purchasing across multiple regions is inconsistent with standard consumer behavior and indicates scalping. \ding{55} & 
    High-frequency attributes (e.g., 80 orders) cannot be judged in isolation. Without cross-referencing user profiles, such patterns may represent legitimate corporate procurement or gifting scenarios. \\ 
    \bottomrule
  \end{tabular}
\end{table*}

While Large Language Models (LLMs) offer a transformative potential to automate labor-intensive investigations, their industrial utility hinges on the seamless integration of expert domain knowledge. In the absence of specialized semantic grounding, general-purpose models often produce spurious reasoning and misaligned judgments. For instance, as illustrated in Table \ref{tab:failure_modes}, an LLM may misidentify concentrated IP clustering in Local Life Service (LLS) as an attack, failing to recognize it as a benign phenomenon inherent to the LLS business. Such failures underscore that domain knowledge is the essential foundation for aligning LLM reasoning with expert domain understanding. 
However, building a knowledge-driven LLM analysis system faces three main challenges.
First, heterogeneous knowledge acquisition is a significant hurdle, as fragmented expertise is often embedded within various sources such as business documentation and codebase. To be effective, this information must be distilled into structured, refined knowledge that provides the analytical depth required for reliable LLM reasoning. 
Second, optimizing contextual knowledge utility remains a primary bottleneck. Unlike standard question-answering systems with explicit queries \cite{lewis2020retrieval}, risk investigation involves complex scenarios where the mapping from case features to relevant domain knowledge is not intuitively defined. This necessitates a tailored design to bridge case contexts with specialized expertise, ensuring that refined knowledge is fully utilized for fine-grained analysis.
Finally, the rapidly shifting e-commerce risk landscape necessitates a self-evolving knowledge-based system. Static knowledge systems lack the capacity to handle new business contexts or new types of attacks.

To address these barriers, we propose \textsc{Sherlock} (\textbf{S}elf-evolving \textbf{H}ierarchy for \textbf{E}-commerce \textbf{R}isk via \textbf{L}LM \textbf{O}rchestrated \textbf{C}ontextual \textbf{K}nowledge), a framework designed to bridge the gap between general LLM capabilities and professional risk investigation requirements. \textsc{Sherlock} is structured around three core functional modules:
\begin{enumerate}[label=\textbf{\arabic*.}, leftmargin=2em]
    \item \textbf{Knowledge Acquisition (How to Build):} We construct a domain Knowledge Base (KB) by combining LLM-based scoring with expert evaluation. We integrate sparse information from multiple sources, such as documentation, code, and meeting minutes, into four distinct knowledge types which provide background knowledge and empirical evidence for model reasoning at various stages. 
    \item \textbf{Knowledge Application (How to Use):} We design a two-stage Retrieval-Augmented Generation (RAG) strategy to optimize the application of domain knowledge. The system uses input augmentation for initial context and a Reflect \& Refine (R\&R) module for logical calibration. The process effectively aligns model judgments with expert investigative standards.
    \item \textbf{Knowledge Evolution (How to Update):} We build a platform integrating expert annotation with daily operations. This platform enables real-time updates of the KB. It also accumulates training samples to support periodic model optimization. This design ensures the system evolves alongside new business scenarios and attack patterns.
\end{enumerate}
The primary contributions of this work are as follows:
\begin{itemize}
    \item We present \textsc{Sherlock}, an industrial-scale framework that integrates LLM reasoning with structured domain knowledge for professional risk investigation.
    \item We develop a pipeline for structured knowledge acquisition and a two-stage RAG strategy to ensure investigative precision and factual alignment.
    \item We introduce a dual-track evolution mechanism for knowledge update and LLM alignment, ensuring the system evolves alongside adversarial tactics.
    \item We provide extensive evaluation from JD.com, demonstrating an 82\% expert acceptance rate and a 386.7\% increase in daily investigation throughput.
\end{itemize}

\begin{figure*}[!tbp]
    \centering
    \includegraphics[width=.95\textwidth]{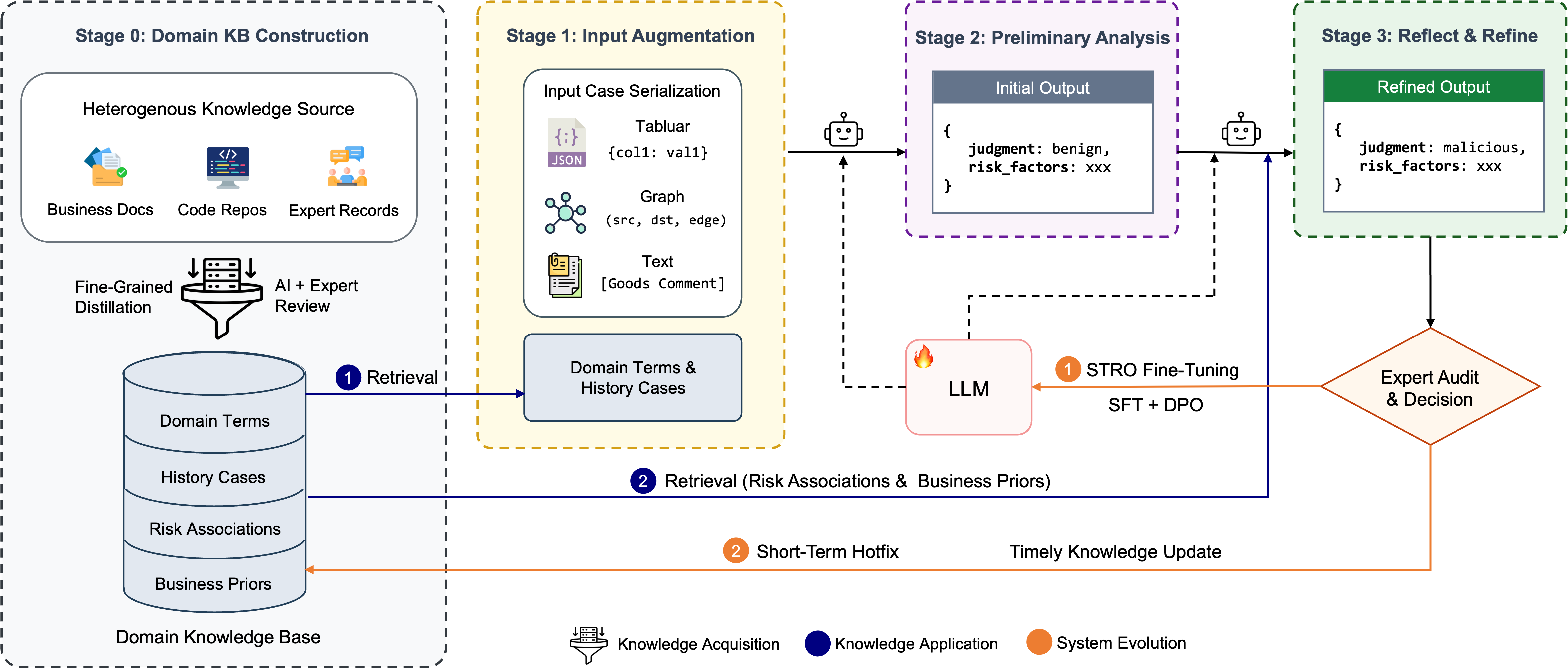}
    \caption{The system architecture of \textsc{Sherlock}. Starting with knowledge acquisition (Stage 0), multi-source data is distilled into the Domain Knowledge Base. Each case follows a sequential pipeline that applies this knowledge through Input Augmentation (Stage 1), Preliminary Analysis (Stage 2), and Reflect \& Refine calibration (Stage 3). Finally, a data flywheel drives continuous system evolution via STRO-based post-training and timely knowledge update.}
    \Description{Architecture diagram of SHERLOCK, featuring a four-stage process from knowledge acquisition, input augmentation, and preliminary analysis to reflect and refine calibration, connected by a data flywheel loop for continuous system updates.}
    \label{fig_pipeline}
\end{figure*}

%% file: sections/related_work.tex
\section{Related Work}
The evolution of e-commerce risk management reflects a necessary transition from manual, rule-based systems to automated, knowledge-driven reasoning \cite{ibrahim2025rule}. Early detection frameworks primarily relied on expert manuals and rigid scoring engines \cite{caron2013comprehensive, song2019cross}, which often struggled to scale or adapt to rapidly shifting adversarial patterns \cite{ali2022financial, fraudant}. The advent of deep learning enabled the capture of complex structural dependencies within transaction networks, particularly through Graph Neural Networks (GNNs) \cite{SAGE, GAT, fraud-recent-3} and hybrid ensemble models \cite{fraud2022deep2, feng2025hybrid}. While methods such as group-based fraud networks \cite{fraud-recent-1} and time-attention frameworks \cite{li2019time, yin2022behavioral} improve detection accuracy, they often function as ``black boxes'' that lack the explainable reasoning chains essential for professional case investigation \cite{gohel2021explainable, rao2020xfraud}.

LLMs such as Llama3 \cite{dubey2024llama} and Qwen3 \cite{yang2025qwen3}, have introduced new possibilities for semantic generalization in e-commerce \cite{e-com-llm-1, li2024ecomgpt, peng2024ecellm}. Recent research has explored enhancing GNNs with LLMs for fraud detection \cite{yang2025flag, huang2025can} and leveraging dual-granularity prompting to improve precision \cite{li2025dgp, luo2025fraud}. Despite these advancements, off-the-shelf models face significant hurdles in industrial settings, notably terminology ambiguity and a lack of grounding in long-tail domain knowledge \cite{huang2025can, liu2025newworldwordembeddings}. While Retrieval-Augmented Generation (RAG) \cite{RAG, e-com-KG} helps mitigate hallucinations \cite{ji2023towards}, current adaptation paradigms like Supervised Fine-Tuning (SFT) \cite{knowledge-inject, ouyang2022training} and Direct Preference Optimization (DPO) \cite{rafailov2023direct, lima} often result in models that are ``frozen'' post-deployment \cite{chen2024general2specialized}. These systems generally lack the agility required to absorb emergent risk patterns in real-time within highly adversarial environments.

To address these limitations, recent studies have begun exploring self-reflection mechanisms \cite{ji2023towards, wang2024taste} and logical reasoning enhancements \cite{logicreasoning, logicllm, luo2025chatrule} to align model outputs with expert intuition. Furthermore, the concept of a self-evolving data flywheel, previously applied in navigation learning \cite{wang2024bootstrapping} and post-training arenas \cite{luo2024arena}, offers a potential path for continuous system optimization. Unlike prior methods that rely on static knowledge structures \cite{e-com-KG, li2025dgp} , \textsc{Sherlock} integrates a dynamic domain knowledge base with a dual-track evolution mechanism to facilitate both immediate risk hotfixes and periodic logic internalization. This establishes a sustainable architecture for high-precision risk detection in dynamic e-commerce scenarios \cite{yang2025fraudr1, yao2025your}.

%% file: sections/problem_formulation.tex
\section{Problem Formulation}
In this section, we formally define the risk investigation task and the objective of dynamic adaptation within the reasoning framework.
\begin{definition}[Risk Investigation Case]
A risk investigation case is defined as $\mathcal{C}$, representing a structured collection of multi-dimensional observations associated with a suspicious entity or transaction. $\mathcal{C}$ encompasses all raw evidence necessary for comprehensive risk assessment, including behavioral records, relational interaction data, and business contexts.
\end{definition}
\begin{definition}[Reasoning and Decision Task]
Given an investigation case $\mathcal{C}$, the objective is to generate an interpretable reasoning trajectory $Y = \{y_1, y_2, \dots, y_n\}$ and a corresponding risk judgment $J \in \{\textit{benign, malicious}\}$. The trajectory $Y$ must explicitly delineate the identified risk signals and the logical transitions that lead to the final decision $J$.
\end{definition}
\begin{definition}[System Objective]
The primary goal is to optimize a reasoning framework $f_\theta$ such that the generated pair $(Y, J)$ maximizes investigative performance and aligns with professional standards. Formally, we seek to optimize:
\[
\arg \max_{\theta} P(Y, J \mid \mathcal{C}; \theta)
\]
where $\theta$ represents the parameters of the underlying model. In the context of highly adversarial e-commerce environments, this objective further requires dynamic evolution. The framework must ensure that the mapping from $\mathcal{C}$ to $(Y, J)$ can be continuously updated or calibrated to maintain high precision as fraud patterns shift or new business scenarios emerge.
\end{definition}

%% file: sections/methods.tex
\section{Methods}
Figure \ref{fig_pipeline} illustrates the overall architecture of \textsc{Sherlock}. The framework consists of three functional modules that form a closed-loop system. First, we develop a pipeline for knowledge acquisition (\textit{how to build}) to construct a domain KB. Second, we implement a multi-stage process for knowledge application (\textit{how to use}) to ground LLM reasoning. Finally, we establish an expert-in-the-loop mechanism for system evolution (\textit{how to update}).

\subsection{Knowledge Acquisition}
\label{sec: domain kb}
This section details the process of distilling reliable and compact knowledge from vast yet low-density industrial data to construct the domain KB.

\subsubsection{Preprocessing of Heterogeneous Knowledge Sources}
We first collect and normalize data from three primary sources into a consolidated repository for downstream extraction. This intermediate repository ensures that heterogeneous information is formatted for efficient parsing while preserving its original semantic depth.

\paragraph{Business Documentation} We utilize a corpus of 1.2 million tokens comprising business SOPs and historical investigation reports. These documents are segmented using a context-aware chunking method to ensure that each unit fits the LLM’s context window while maintaining semantic continuity.

\paragraph{Code and Strategies} We target feature calculation code and risk discrimination models within the platform’s code repositories. Preprocessing involves parsing conditional logic and mapping symbolic variables to their natural language descriptions (e.g., ``ord\_cnt\_1w'' is translated to ``weekly order count'').

\paragraph{Expert Records} We process unstructured materials, including expert meeting transcripts and case-specific review notes. To reduce noise, we implement a filtering step to remove administrative overhead and non-essential dialogue, ensuring that the preprocessing focuses on the segments that record the expert's decision-making logic and investigative reasoning.

\subsubsection{Fine-Grained Knowledge Distillation and Classification}
From the preprocessed data, we extract four types of structured knowledge. This classification is designed to support the two-stage RAG strategy detailed in Section \ref{subsec: rag} to maximize the utility of the KB. To balance knowledge coverage with annotation costs, we use an LLM to extract initial knowledge candidates, which are then reviewed and refined by domain experts. For reproducibility, the specific prompt template are provided in Appendix \ref{app:prompts}.

\paragraph{Domain Terminology} An LLM-based self-checking mechanism identifies terms where intrinsic model knowledge deviates from specialized business definitions. Experts then certify these definitions to align the LLM’s vocabulary with platform-specific jargon.

\paragraph{History Cases} Raw case logs and notes are transformed into structured reference exemplars containing the case description, the final judgment, and the underlying expert reasoning. These serve as historical anchors for analogical reasoning during inference.

\paragraph{Risk Associations} We extract the inter-dependencies between distinct risk factors. These associations are derived from the semantic context within business documentation and logical conjunctions (e.g., ``AND'' operators) found in the codebase. This category defines how multiple risk factors combine to signal a potential attack. Experts verify these connections to ensure they represent actual adversarial logic rather than isolated observations.

\paragraph{Business Priors} We extract plausible justifications for specific behavioral patterns using risk factors as anchors. Senior operators validate these logic entries to ensure they effectively whitelist benign behaviors and reduce false positives in production.

\subsection{Knowledge-Augmented Inference}
\label{subsec: rag}
This section describes the integration of the KB into the inference phase via a two-stage RAG process: input augmentation for initial context and the R\&R module for logical calibration.

\subsubsection{Input Augmentation}
The initial stage anchors the model’s reasoning in domain context through targeted retrieval. We first serialize the input case into a structured Markdown format. To resolve specialized jargon, we utilize an LLM-based self-checking mechanism to extract terms with potential ambiguity as queries for retrieving \textit{domain terminology}. Simultaneously, the system retrieves top-$k$ \textit{history cases} from the KB based on the semantic similarity of the case description. These retrieved elements are integrated into the input, ensuring the preliminary analysis aligns with platform-specific definitions and historical evidence.

\subsubsection{Reflect \& Refine}
The R\&R module serves as a calibration engine for the model's initial findings. It utilizes the preliminary risk factors identified in the first stage as search keys to trigger a second-pass retrieval. This targeted search specifically fetches \textit{risk associations} and \textit{business priors} from the KB. The module then performs three core verification tasks:
\textbf{(1) Factual Verification.} This task compares the initial risk factors with the original input data. It checks for factual errors to ensure the model correctly interpreted the heterogeneous features (e.g., transaction counts or graph connections). The goal is to identify and fix misinterpretations that may have occurred during the first-pass analysis. \textbf{(2) Risk Association.} This task uses the retrieved \textit{risk associations} to identify potential threats that were missed in the initial stage. This step ensures a more comprehensive analysis by uncovering hidden risks not immediately caught by the model's general reasoning. \textbf{(3) Business Justification.} This task uses the retrieved \textit{business priors} to find legitimate explanations for suspicious activities. It aims to reduce false positives by checking if flagged behaviors match valid business scenarios.

\begin{figure*}[ht]
    \centering
    \includegraphics[width=.8\textwidth]{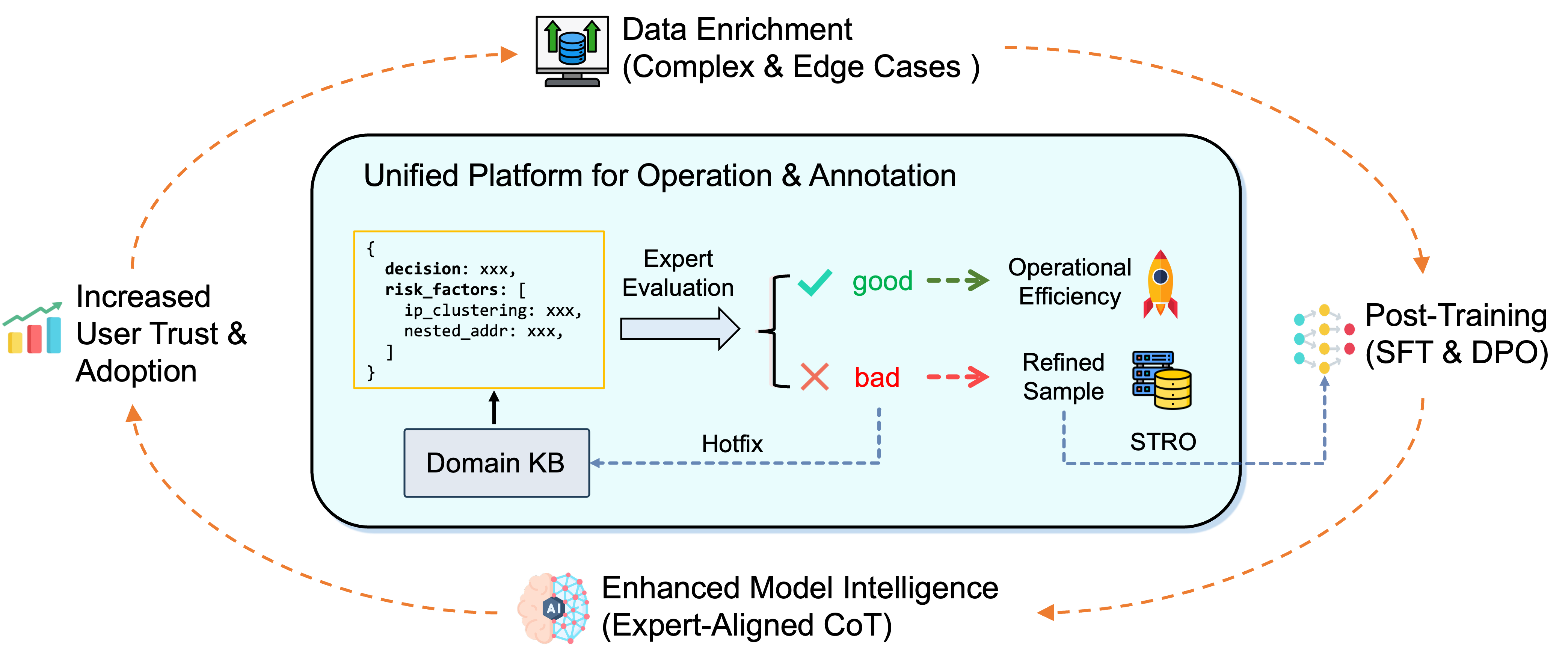}
    \caption{The data flywheel of \textsc{Sherlock}, integrating expert-in-the-loop operations with continuous system evolution.}
    \Description{A circular flowchart illustrating the data flywheel mechanism of the Sherlock framework, showcasing the interaction loop between human expert evaluation, online knowledge base hotfixes, and continuous offline post-training alignment.}
    \label{fig_dataflywheel}
\end{figure*}

\subsection{Knowledge-Driven Model Alignment}
We align the model with expert logic through a knowledge-driven fine-tuning process. This process leverages the domain KB and the R\&R module to construct preference datasets. To ensure reasoning quality, the \textit{Suspect-Then-Rule-Out} (STRO) framework guides the generation of reliable CoT sequences for model optimization.

\subsubsection{Expert-Aligned Trajectory Synthesis}
To transform discrete expert feedback into high-quality supervision signals that capture professional reasoning, we develop a trajectory synthesis pipeline called STRO based on expert review results.
STRO emulates professional investigative patterns through a two-phase CoT process. The model first performs a broad identification of potential risk factors from the input case, mirroring the preliminary results in Stage 2 shown in Figure \ref{fig_pipeline}. It then executes a critical self-reflection step: using logical verification and expert experience from the KB, the model rules out incorrect risks and incorporates missing ones. By using expert-validated factors as anchors, STRO transforms simple labels into high-quality reasoning paths that arrive at the final expert-approved conclusion. This ensures the data captures the underlying logical transitions required for complex risk assessment. The prompt template for this process is provided in Appendix \ref{app:stro_prompt}.

\subsubsection{Preference Alignment}
To internalize these patterns, we implement a two-stage alignment strategy:

\paragraph{Supervised Fine-Tuning (SFT)} We first perform SFT using expert-verified STRO trajectories $\mathcal{D}_{sft} = \{(x^{(i)}, y^{(i)})\}_{i=1}^N$. The objective is to minimize the standard cross-entropy loss, ensuring the model learns the basic structure of expert reasoning:
\begin{equation*}
\mathcal{L}_{SFT}(\theta) = -\mathbb{E}_{(x,y) \sim \mathcal{D}_{sft}} \sum_{t=1}^{|y|} \log P_{\theta}(y_t | x, y_{<t})
\end{equation*}

\paragraph{Direct Preference Optimization (DPO)} In e-commerce risk management, minimizing false positives is critical, as misjudgments can lead to significant financial loss and adverse user sentiment. To address this, we utilize DPO to further distinguish between subtle hallucinations and genuine risk signals. We construct the preference triplets $(x, y_w, y_l)$ leveraging the outputs from different stages of our pipeline. Specifically, the \textbf{winning sample} ($y_w$) is the expert-verified conclusion with an STRO reasoning trajectory, while the \textbf{losing sample} ($y_l$) is the corresponding output from the preliminary analysis stage.
By optimizing the model to prefer $y_w$ over $y_l$, we encourage it to move beyond superficial anomalies toward evidence-based conclusions.
For each triplet $(x, y_w, y_l) \in \mathcal{D}_{dpo}$, the objective is to minimize:
\begin{equation*}
\mathcal{L}_{DPO}(\theta; \pi_{ref}) = -\mathbb{E}_{(x, y_w, y_l) \sim \mathcal{D}_{dpo}} \big[ \log \sigma \big(h_\theta(y_w, x) - h_\theta(y_l, x) \big) \big]
\end{equation*}
where $h_\theta(y, x) = \beta \log \frac{\pi_\theta(y|x)}{\pi_{ref}(y|x)}$, $\pi_{ref}$ is the model initialized from the SFT stage and $\beta$ is a hyperparameter controlling the strength of the KL penalty.

\subsection{Efficient Knowledge Evolution}
\label{subsec: data flywheel}
\textsc{Sherlock} is designed as an evolvable system that improves through a continuous feedback loop, see Figure \ref{fig_dataflywheel}. As the model aligns with expert reasoning via the post-training pipeline, its analytical accuracy increases, leading to lower false-positive rates and greater operational trust. This performance gain drives broader adoption across teams, which in turn surfaces a high volume of complex edge cases and expert interactions. These data points are captured to further refine the training process, creating a self-reinforcing data flywheel.

\subsubsection{Unified Platform}
The core of our engineering implementation is a unified platform designed to eliminate the bottleneck of human labor in risk management. As shown in Figure \ref{fig_dataflywheel}, we achieve a dual-efficiency gain by consolidating daily operations and data labeling into a single interface. \textbf{(1) Operational Efficiency}: The platform presents LLM judgment and risk factors to senior experts during routine review. The time required for case investigation is significantly reduced, allowing experts to handle a higher volume of complex cases with less effort. \textbf{(2) Annotation Efficiency}: To ensure continuous improvement, experts provide brief corrections for suboptimal or erroneous model responses during daily operations. The STRO framework then uses these corrections to automatically construct high-quality training samples. This process significantly reduces the annotation workload by eliminating the need for independent and large-scale manual labeling sessions. 

\subsubsection{Dual-track Evolution: Hotfix and Training}
To maintain reliability in the volatile e-commerce environment, we implement a complementary update strategy that addresses both short-term urgency and long-term robustness:

\begin{itemize}
    \item \textbf{Short-term Hotfix (Timely Knowledge Update)}: For emerging fraud tactics, the domain KB and R\&R module provide a ``hotfix'' capability. Operators can inject new knowledge, such as the definition of a new business scenario and the description of an emerging tactic to immediately update the system's analytical capabilities with near-zero latency.
    \item \textbf{Logic Internalization (SFT + DPO)}: While hotfixes offer rapid intervention, long-term stability is achieved through logic internalization. Periodic post-training ensures accumulated insights are structurally integrated into the model's core reasoning.
\end{itemize}

%% file: sections/experiments.tex
\section{Experiments}
This section details the empirical evaluation of our proposed method. We aim to answer the following Research Questions (RQs):
\begin{enumerate}[label=\textbf{\arabic*.}, leftmargin=2em]
    \item \textbf{RQ1 (Overall  Effectiveness):} How does the performance of \textsc{Sherlock}-enhanced small-scale models compare to that of massive foundation models in specialized risk investigation?
    \item \textbf{RQ2 (Ablation Study): } What are the marginal contributions of the domain KB, R\&R module, and STRO framework to reasoning precision and hallucination suppression?
    \item \textbf{RQ3 (Industrial Impact): } What is the operational performance of \textsc{Sherlock} in real-world scenarios, and how does the data flywheel facilitate autonomous system evolution to counteract adversarial drift?
\end{enumerate}

\subsection{Experimental Setup}
\subsubsection{Dataset}
Experiments are conducted on an anonymized real-world dataset of 5,000 cases from JD.com, collected over a one-month period via our unified platform. To ensure semantic integrity, heterogeneous features comprising tabular attributes, graph-based relationships, and unstructured text are unified into a structured Markdown format. Using the STRO framework, we synthesize high-fidelity reasoning trajectories: reasoning paths are first generated by \textit{JoyAI-750B} with STRO, and subsequently refined by a panel of 15 senior risk experts to eliminate hallucinations and ensure strict alignment with verified business logic. This expert-verified corpus provides the high-quality supervision signals necessary for the subsequent supervised fine-tuning and preference optimization stages.

\begin{table*}[ht]
  \caption{Overall model performance comparison on the experimental benchmark.}
  \label{tab: overall evaluation}
  \begin{tabular}{cc|ccc|cccc}
    \toprule
    Method & Model & FAR & SNR & CDR & Precision & Recall & $F_1$ & FPR\\
    \midrule
    Legacy & Hybrid & \textemdash & \textemdash & \textemdash & \textbf{0.95} & 0.82 & \textbf{0.88} & \textbf{0.04} \\
    \midrule
    \multirow{4}{*}{Vanilla} & DeepSeek-R1 & 0.75 & 0.73 & 0.58 & 0.52 & 0.74 & 0.61 & 0.68 \\
    & JoyAI-750B & 0.78 & 1.03 & 0.56 & 0.55 & 0.71 & 0.62 & 0.58 \\
    & JoyAI-10B & 0.63 & 0.86 & 0.41 & 0.42 & 0.65 & 0.51 & 0.90\\
    & Qwen3-8B & 0.65 & 0.79 & 0.45 & 0.45 & 0.68 & 0.54 & 0.83 \\
    \midrule
    \multirow{2}{*}{\textsc{Sherlock}} & JoyAI-10B & \textbf{0.92} & \textbf{4.34} & \textbf{0.64} & 0.89 & 0.84 & 0.86 & 0.10 \\
    & Qwen3-8B & 0.89 & 4.26 & \textbf{0.64} & 0.85 & \textbf{0.86} & 0.85 & 0.15 \\
    \bottomrule
  \end{tabular}
\end{table*}

\subsubsection{Domain KB}
A core component of our framework is the domain KB. The KB was constructed by collaborating with 15 senior risk experts. At the time of initialization, the repository contains in total 1850 entries formatted as structured JSON objects, which consists of 1408 domain terms, 350 history cases, 52 risk associations and 40 business priors. The KB is indexed using a hybrid keyword and vector search system (BGE-M3 \cite{chen2024bge}) to ensure robust retrieval.

\subsubsection{Models and Methods}
We evaluate a comprehensive set of baselines and our proposed methods to demonstrate a clear progression of capabilities. \textit{JoyAI-10B} and \textit{Qwen3-8B} are selected as base models for fine-tuning due to their domain-specific pre-training on e-commerce logic.

\paragraph{Models Under Comparison} We compare two tiers of model capacity: (1) Foundation Models, including our in-house \textit{JoyAI-750B} and \textit{DeepSeek-R1}\cite{deepseek-r1}; and (2) Base Models, specifically \textit{JoyAI-10B} and \textit{Qwen3-8B}, used as backbones for domain-specific internalization.

\paragraph{Methods} We contrast three distinct method configurations: (1) \textit{Legacy}, the traditional non-LLM hybrid system (including rule engines, XGBoost and GNN detectors) currently used in production; (2) \textit{Vanilla}, which performs zero-shot inference using the foundation or base models without any domain-specific adaptation; and (3) \textsc{Sherlock}, our end-to-end framework that integrates the STRO-fine-tuned base model with the domain KB and R\&R module.

\subsubsection{Evaluation Benchmarks}
To provide a holistic assessment of the \textsc{Sherlock} framework, we evaluate both the fine-grained reasoning trajectories and the final risk classification accuracy.
We curated an evaluation set of 1,000 samples. To ensure high-quality benchmarking, our panel of 15 senior risk experts performed detailed annotation of each case, identifying \textit{core risk factors} (decisive for the final judgment) and \textit{relevant risk factors} (supportive context required for investigation), as well as the final decision. To streamline the evaluation pipeline, we utilized an LLM-as-a-Judge to categorize generated factors based on expert annotations, while the final decision was evaluated through direct string matching against the expert-certified labels.
The following metrics are designed to assess the quality of the generated risk factors:
\begin{itemize}
    \item Factual Alignment Rate (FAR): The ratio of generated risk factors that accurately reflect input facts, quantifying the model's data comprehension and hallucination frequency.
    \item Signal-to-Noise Ratio (SNR): The ratio of the total number of core and relevant risk factors to the number of noise instances, measuring the refinement level of the model’s output  (the most critical metric we focus on).
    \item Core-risk Detection Rate (CDR): The proportion of core risk factors recalled by model. This metric is crucial for evaluating the ability of the model to capture key risk factors.
\end{itemize}
In addition, to directly assess the model's performance in distinguishing \textit{malicious} from \textit{benign} transactions, we report standard \textit{Precision}, \textit{Recall}, and $F_1$ score, treating \textit{malicious} cases as the positive class.

\subsection{Overall Performance}
We evaluate the effectiveness of the \textsc{Sherlock} framework by comparing its performance against various baselines across reasoning quality and judgment accuracy, as summarized in Table \ref{tab: overall evaluation}.

Compared to the legacy \textit{Hybrid} system, \textsc{Sherlock} offers two distinct industrial advantages. First, while the legacy system operates as a ``black box'', \textsc{Sherlock} generates structured reasoning trajectories that transform the expert's role from ``data miner'' to ``fact verifier'', significantly reducing investigation time. Second, \textsc{Sherlock} achieves superior \textit{Recall} (up to 0.86 vs. 0.82), as its semantic generalization captures long-tail adversarial patterns that rigid rules, characterized by high \textit{Precision}, often fail to detect. 

Notably, \textsc{Sherlock} enables 10B-scale models to surpass significantly larger foundation models. While \textit{JoyAI-750B} achieves an $F_1$ score of 0.62, \textit{JoyAI-10B} enhanced with \textsc{Sherlock} reaches 0.86. This 38.7\% relative improvement validates that structured expert reasoning and factual guardrails outweigh the advantages of raw model parameter scale in specialized risk domains.

Vanilla models are susceptible to ``risk hallucinations'' in which statistical noise is misinterpreted as definitive evidence, ultimately leading to excessive false positives. \textsc{Sherlock} effectively mitigates this by prioritizing verifiable evidence , increasing the \textit{Precision} for the \textit{JoyAI-10B} backbone from 0.42 to 0.89.

The consistent performance gains across both \textit{JoyAI} and \textit{Qwen} backbones validate that \textsc{Sherlock} is not overfitted to a specific architecture. This model-agnostic nature demonstrates the framework's broad reusability and its potential for seamless integration into diverse industrial LLM ecosystems.

\subsection{Ablation Study}
In this section, we investigate the marginal contribution of each core component within \textsc{Sherlock}. To isolate the effects of the reasoning loop and external knowledge, we evaluate three variants against the full framework: (1) w/o R\&R, which uses a single-pass inference with KB augmentation; (2) w/o KB, which maintains the R\&R loop but relies solely on the model's internal parametric memory; and (3) w/o STRO, which is a standard fine-tuned baseline. Table \ref{tab:ablation} summarizes the performance of these variants on the \textit{JoyAI-10B} backbone.

\begin{table}[ht]
\centering
\caption{Quantifying the contribution of core components in \textsc{Sherlock} built upon the JoyAI-10B backbone.}
\label{tab:ablation}
\begin{tabular}{l|ccc|c}
\toprule
Variant & FAR & SNR & CDR & $F_1$ \\
\midrule
\textsc{Sherlock} (Full) & \textbf{0.92} & \textbf{4.34} & 0.64 & \textbf{0.86} \\
\midrule
w/o R\&R & 0.83 & 2.98 & \textbf{0.67} & 0.84 \\
w/o KB  & 0.80 & 1.98 & 0.59 & 0.79 \\
w/o STRO & 0.81 & 2.53 & 0.63 & 0.75 \\
\bottomrule
\end{tabular}
\end{table}

As shown in Table \ref{tab:ablation}, removing the R\&R loop leads to a substantial degradation in reasoning precision, with the SNR falling from 4.34 to 2.98 and the FAR dropping from 0.92 to 0.83. Although this variant maintains a higher CDR (0.67), the collapse in SNR confirms that without the reflection and refinement loop, the model reverts to an ``aggressive'' detection mode. The R\&R module thus functions as a critical filter that utilizes domain expertise to distinguish actual risk from non-essential business noise. Additionally, the removal of the domain KB results in the most severe collapse of reasoning quality, with the SNR falling from 4.34 to 1.98. This 54\% reduction in SNR, coupled with a drop in $F_1$ to 0.79, demonstrates that internal parametric memory alone cannot sustain professional-grade investigation. Without external domain expertise, specifically risk associations and specialized business logic, the model loses its ability to distinguish core risk signals from background noise. The decline in FAR (0.92 to 0.80) further proves that this expertise is the primary driver for maintaining logical grounding and preventing the model from wandering into plausible but irrelevant reasoning paths.
Moreover, the \textit{w/o STRO} variant yields the lowest overall $F_1$ score (0.75), confirming that the ``\textit{Suspect-Then-Rule-Out}'' fine-tuning is the indispensable prerequisite for the entire system. While it maintains a slightly higher SNR (2.53) than the \textit{w/o KB} variant, the comprehensive decline across all metrics indicates that without this specialized cognitive alignment, the model fails to properly leverage the domain KB or the R\&R module. This validates that the STRO stage provides the fundamental reasoning structure required to initiate and sustain the complex investigative logic that \textsc{Sherlock} targets.

\subsection{Practical Deployment and Business Impact}
\begin{table*}[t] 
\centering 
\caption{Online performance of the initial \textsc{Sherlock} system during A/B test.} 
\label{tab: online exp} 
\begin{tabular}{l|cc|c} 
\toprule 
\textbf{Metrics} & \textbf{Legacy (Control)} & \textbf{\textsc{Sherlock} (Treatment)} & \textbf{Improvement} \\ 
\midrule 
Avg. Review Time per Case (min) & 15.24 & 3.13 & \textbf{-79.5\%} \\ 
P95 Review Time (min) & 45.10 & 8.50 & \textbf{-81.2\%} \\ 
Expert Acceptance Rate & 0.31 & 0.82 & \textbf{+164.5\%} \\ 
Avg. Daily Throughput (case) & 945 & 4600 & \textbf{+386.7\%} \\ 
\bottomrule 
\end{tabular} 
\end{table*}
\textsc{Sherlock} was deployed in the production environment at JD.com to quantify its real-world impact through a randomized A/B experiment. We implemented an $80/20$ traffic split, where 80\% of the live requests were routed to the treatment group utilizing \textsc{Sherlock}'s analysis pipeline, while the remaining 20\% served as the control group under the legacy rule-based system. In this control group, investigative support was provided by a template-driven engine that generated standardized textual justifications by mapping the top-ranking feature importance from the underlying hybrid models, including rule engines, GNNs, and tree-based detectors.

As shown in Table \ref{tab: online exp}, \textsc{Sherlock} significantly outperformed the legacy system. The \textit{average review time per case} decreased from 15.24 minutes to 3.13 minutes. This 79.5\% efficiency gain is primarily attributed to \textsc{Sherlock}’s ability to synthesize heterogeneous data into a human-readable \textit{reasoning trajectory}. Moreover, the \textit{Expert Acceptance Rate} (EAR) rose from 31\% to 82\%, demonstrating that the LLM's reasoning aligns closely with high-level expert intuition.
A critical highlight for industrial deployment is the reduction in \textit{P95 review time}. Complex and multi-layered fraud cases often cause significant operational bottlenecks in risk management. \textsc{Sherlock} reduced the P95 latency from 45.10 minutes to just 8.50 minutes (-81.2\%). This indicates that even for the most intricate 5\% of cases, the framework provides robust reasoning support, ensuring consistent operational throughput during high-pressure periods.

\textsc{Sherlock} significantly expands investigative capacity by removing the throughput bottlenecks inherent in manual review. As shown in Table \ref{tab: online exp}, while the legacy system's capacity was capped at 945 cases per day, \textsc{Sherlock} increased daily throughput to 4,600 cases—a $386.7\%$ improvement. This allows the platform to maintain broad risk coverage without proportional increases in headcount.

\subsection{Continuous Improvement via Data Flywheel}
\begin{figure}
    \centering
    \includegraphics[width=\linewidth]{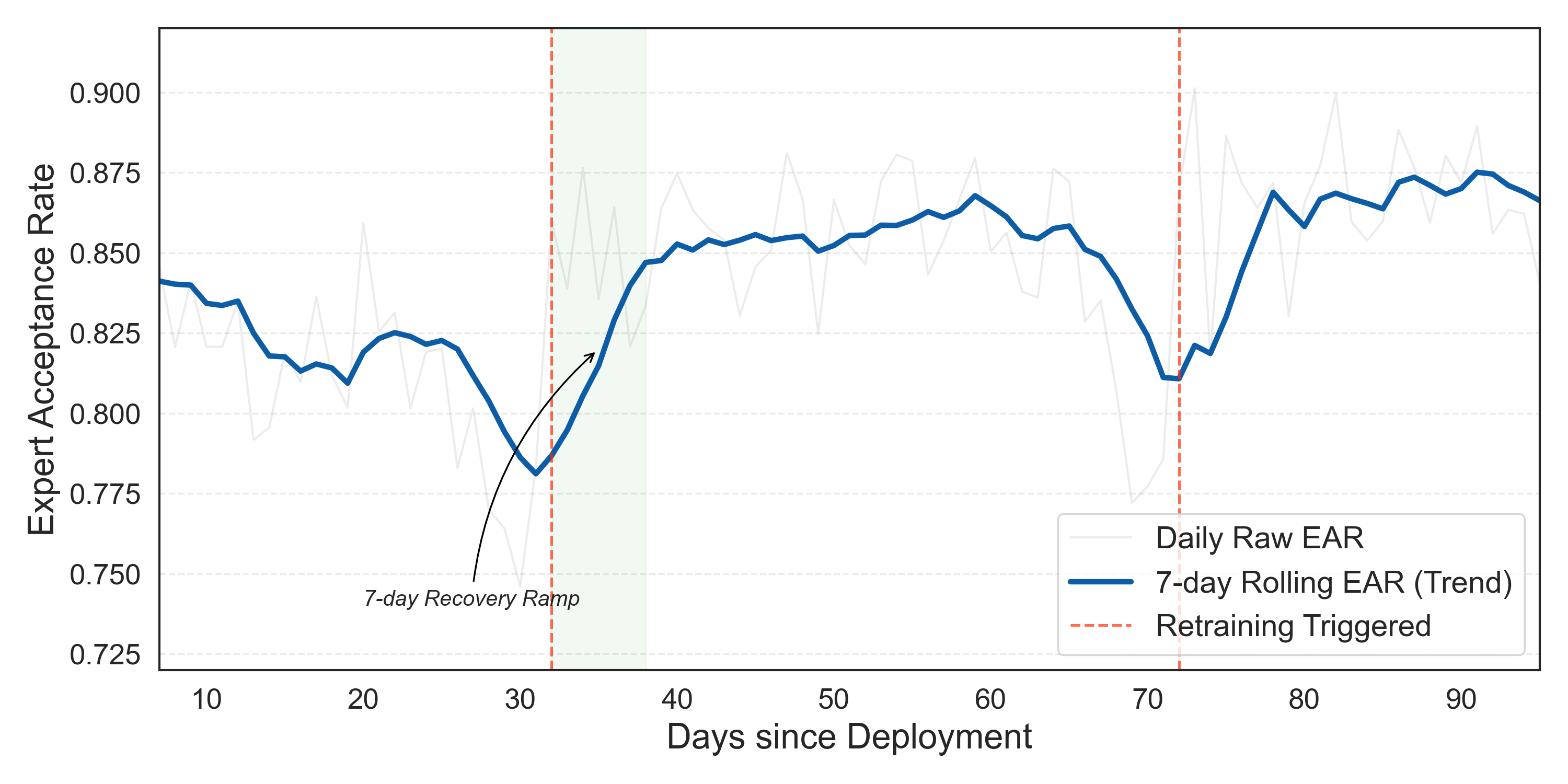}
    \caption{Self-evolving performance via autonomous data flywheel over a 90-day deployment period.}
    \Description{A performance evaluation line graph plotting the Expert Acceptance Rate against 90 days of deployment. It highlights a 7-day rolling trend curve along with two vertical indicators marking when model retraining was triggered, showing a clear performance drop before and recovery after each trigger.}
    \label{fig_dataflywheel_exp}
\end{figure}
To ensure the long-term sustainability of \textsc{Sherlock} in a dynamic adversarial environment, we implement an autonomous data flywheel mechanism. This design facilitates a self-evolving loop where high-quality annotations and reasoning corrections from the unified platform are back-propagated to iteratively refine the system.

\paragraph{Monitoring and Trigger Mechanism} We employ the 7-day rolling EAR as the primary health indicator for the production model. By utilizing a 7-day sliding window, we effectively smooth out daily operational fluctuations while maintaining sensitivity to genuine distribution shifts. Our system is configured with an automated retraining trigger: whenever the rolling EAR exhibits a downward trend and drops by 5\% relative to stable performance following the previous model update, a new fine-tuning cycle is initiated.

\paragraph{Observations of Autonomous Evolution} Over a 90-day observation period, the performance trajectory reveals a distinctive ``Decay-Ramp-Stabilize'' pattern: 
\textbf{(1) The Recovery Ramp.} A critical observation in Figure \ref{fig_dataflywheel_exp} is that following a retraining trigger (e.g., at Day 32), the rolling EAR exhibits a progressive upward slope spanning approximately 7 days. This progressive upward trend validates that model updates, fueled by the ingestion of new expert-annotated cases, effectively enhance the system's operational capabilities. \textbf{(2) The Ratchet Effect.} Each iteration of the flywheel not only restores the EAR but establishes a new, higher performance ceiling. As shown in Figure \ref{fig_dataflywheel_exp}, the EAR stabilized at 86\% after the first cycle and reached a new peak of 87\% after the second, demonstrating the framework's ability to consolidate new adversarial insights without ``catastrophic forgetting''.

This feedback loop enables \textsc{Sherlock} to adapt to new fraud patterns automatically, eliminating the need for manual heuristic updates. 

%% file: sections/conclusion.tex
\section{Conclusion}
This paper introduces \textsc{Sherlock}, a framework that systematically overcomes the primary challenges of knowledge acquisition, application, and evolution in e-commerce risk management. We demonstrate that distilling structured expertise from low-density industrial sources provides the necessary semantic grounding to align LLMs with specialized business logic. To ensure effective knowledge application, the two-stage RAG strategy and R\&R module prove that iterative logic calibration can successfully eliminate hallucinations and noise. Finally, the data flywheel architecture addresses knowledge evolution by establishing a sustainable dual-track mechanism; this allows the system to absorb emerging fraud patterns through autonomous updates, removing the traditional reliance on manual heuristic maintenance.

The industrial efficacy of \textsc{Sherlock} is confirmed by its production deployment at JD.com, where it significantly improved both precision and operational capacity. Online A/B tests resulted in an 82\% EAR and a 386.7\% increase in daily investigation throughput, demonstrating that the framework handles high-volume traffic without requiring proportional increases in headcount. Furthermore, a 90-day production evaluation confirms the robustness of the system's self-evolution; the data flywheel successfully recovered from performance decay on two separate occasions and raised the EAR ceiling by approximately 3.5\% through autonomous post-training. While validated in e-commerce, the architectural paradigm of \textsc{Sherlock} is inherently domain-agnostic and readily generalizable to other high-precision, adversarial decision-making fields.

Despite these gains, certain limitations offer directions for future work. While the framework improves efficiency, expanding the scope of automated knowledge extraction remains a priority to better cover highly complex edge cases. Additionally, reducing the computational overhead and latency associated with multi-stage reasoning presents an opportunity for further optimization. Future research will focus on refining the knowledge distillation pipeline and optimizing the reasoning engine to support more granular, real-time joint investigations across diverse industrial domains.

%% file: sections/appendix.tex
\section{LLM Prompt Framework and Detailed Instructions}
\label{app:prompts}

To ensure the reproducibility of our knowledge distillation process, we provide the complete prompt framework. This includes a \textit{Generic Template} that acts as a global wrapper and four \textit{Detailed Instructions} tailored to each knowledge category.

\subsection{Generic Knowledge Distillation Template}
\label{subsec:generic_template}
The following template defines the expert role and the structural constraints. The placeholders are dynamically populated by the specific instructions provided in the subsequent section.

\begin{promptbox}{Unified Generic Knowledge Distillation Template}
\label{lst:generic_prompt}
\textbf{\# Role} \\
You are a Senior E-commerce Risk Management Expert. Your goal is to distill high-density domain knowledge from the provided preprocessed industrial data.\\

\textbf{\# Input Information} \\
- Source Type: \{Documentation | Code | Expert Record\} \\
- Raw Content: \{\{Preprocessed\_Content\_Snippet\}\} \\
- Risk Factor Pool: \{\{Pre-defined\_Risk\_Factors\_Set\}\} \\

\textbf{\# Task-Specific Instruction} \\
\{\{TASK\_INSTRUCTION\}\} \\

\textbf{\# Output Format \& Constraints} \\
- Return the result strictly in the following JSON schema: \\
\{\{JSON\_SCHEMA\}\} \\
- Do not include any preamble, administrative noise, or non-essential dialogue. \\
- Ensure all extracted information is grounded in the provided Raw Content. \\
- For Association/Priors, only extract data related to the provided Risk Factor Pool.
\end{promptbox}

\subsection{Detailed Task-Specific Instructions}
Following the framework in \ref{lst:generic_prompt}, we provide four specialized instruction sets tailored to generate prompts for distinct types of risk domain knowledge.

\begin{promptbox}{Instruction A: Domain Terminology (Vocabulary Alignment)}
\textbf{\# TASK\_INSTRUCTION:}
\begin{enumerate}[leftmargin=1.5em, label=\arabic*., noitemsep]
    \item \textbf{Identification:} Scan the provided content for platform-specific jargon, acronyms, or internal business concepts.
    \item \textbf{Alignment Check:} Determine if the term's meaning in this industrial context deviates from general-purpose language or standard LLM pre-training knowledge.
    \item \textbf{Expert Definition:} Generate a concise definition that captures the specific business logic or operational utility of the term as described in the content.
\end{enumerate}

\textbf{\# JSON\_SCHEMA:}
\begin{lstlisting}[style=jsonstyle]
{
  "term": "The identified jargon or acronym",
  "definition": "The expert-certified platform-specific definition"
}
\end{lstlisting}
\end{promptbox}

\begin{promptbox}{Instruction B: History Cases (Reasoning Anchors)}
\textbf{\# TASK\_INSTRUCTION:}
\begin{enumerate}[leftmargin=1.5em, label=\arabic*., noitemsep]
    \item \textbf{Case Synthesis:} Condense unstructured investigation logs into a structured summary of user behavior and data anomalies.
    \item \textbf{Outcome Attribution:} Identify the expert's final judgment as either ``Malicious'' or ``Benign''.
    \item \textbf{Evidence Chain Mapping:} Trace the causal links. Why did the expert reach this conclusion? Extract key features that served as the "smoking gun" in the investigation.
\end{enumerate}

\textbf{\# JSON\_SCHEMA:}
\begin{lstlisting}[style=jsonstyle]
{
  "case_id": "Unique identifier (e.g., CASE_001)",
  "description": "Summary of behavioral patterns and data points",
  "judgment": "Benign | Malicious",
  "rationale": "Step-by-step expert reasoning path and evidence chain"
}
\end{lstlisting}
\end{promptbox}

\begin{promptbox}{Instruction C: Risk Associations (Logical Inter-dependencies)} 
\textbf{\# TASK\_INSTRUCTION:} 
\begin{enumerate}[leftmargin=1.5em, label=\arabic*., noitemsep] 
    \item \textbf{Factor Grounding:} Identify and locate occurrences of risk factors from the \textit{Risk Factor Pool} within the provided domain knowledge and logical context. 
    \item \textbf{Dependency Analysis:} Extract the logical inter-dependencies (e.g., conditional sequences, causal links, or Boolean operators) that govern the interactions between these factors. 
    \item \textbf{Graph Construction:} Map the structural relationships between primary and associated risks to characterize coordinated adversarial attack patterns.
\end{enumerate} 
\end{promptbox}

\begin{promptbox}{Instruction D: Business Priors (Benign Justifications)}
\textbf{\# TASK\_INSTRUCTION:}
\begin{enumerate}[leftmargin=1.5em, label=\arabic*., noitemsep]
    \item \textbf{FP Sensitivity Check:} Identify risk factors from the \textit{Risk Factor Pool} that often trigger false positives in specific business scenarios.
    \item \textbf{Contextual Validation:} Extract the specific business context (e.g., seasonal promotional events, special scenarios) where the anomalous behavior is actually legitimate.
    \item \textbf{Logic Formulation:} Formulate a whitelisting justification that justifies why this specific pattern should be considered benign in the identified context.
\end{enumerate}

\textbf{\# JSON\_SCHEMA:}
\begin{lstlisting}[style=jsonstyle]
{
  "risk_factor": "The factor prone to false positives",
  "business_logic": "The detailed justification for whitelisting or benign labeling"
}
\end{lstlisting}
\end{promptbox}

\subsection{Prompt for STRO Reasoning Generation}
\label{app:stro_prompt}

The \textit{Suspect-Then-Rule-Out} (STRO) framework is employed in the construction of training data for model alignment. It aims to reconstruct the complete expert reasoning path by contrasting preliminary suspicions with expert-validated conclusions.

\begin{promptbox}{Prompt of Suspect-then-Rule-Out}
\textbf{\# TASK DESCRIPTION} \\
You are a senior e-commerce risk management expert. Your task is to ``reverse-engineer'' and ``recreate'' your initial, complete internal thought process as the expert, based on the final review conclusion (which risks were accepted and which were rejected). This process must be presented as a first-person soliloquy and must sound like a genuine expert analyzing data in real-time.

\medskip
\textbf{\# INPUT INFORMATION}
\begin{enumerate}[leftmargin=1.5em, label=\arabic*., noitemsep]
    \item \textbf{input data}: Raw input data.
    \item \textbf{accepted risks}: List of risk factors you ultimately confirmed and accepted.
    \item \textbf{rejected risks}: List of risk factors you initially suspected but ultimately ruled out.
\end{enumerate}

\medskip
\textbf{\# CORE REQUIREMENTS FOR THE SOLILOQUY} \\
Your soliloquy must be a coherent narrative that skillfully integrates the following two thought processes:
\begin{itemize}[leftmargin=1.5em, noitemsep]
    \item \textbf{1. For Accepted risk factors (Confirmation Logic):}
    \begin{itemize}[label=--]
        \item Describe how you identified the relevant anomalous data.
        \item Demonstrate how you connected the clues and applied risk management knowledge to confirm their riskiness.
        \item Use affirmative, confident language (e.g., ``This point is suspicious,'' ``This confirms my suspicion'').
    \end{itemize}
    \item \textbf{2. For Rejected risk factors (Exclusion Logic):} \\
    Must reflect the ``suspect first, rule out later'' process.
    \begin{itemize}[label=--]
        \item \textbf{Step 1: Acknowledge surface anomaly.} First, acknowledge why this point ``appeared'' to be a risk (e.g., ``When I first saw this huge amount of orders, I thought it was abnormal...'').
        \item \textbf{Step 2: Seek plausible explanation.} Describe your process of seeking more contextual information to validate this suspicion (e.g., ``But I won't jump to conclusions yet; let me look at this user's background...'').
        \item \textbf{Step 3: Find exclusion evidence.} Clearly identify the specific data point or business knowledge that led you to dismiss the initial suspicion (e.g., ``Ah, I see the user segment is `Verified Enterprise Customer.' This amount is routine for their procurement, so that's fine.'').
        \item \textbf{Step 4: State exclusion conclusion.} Clearly state that you have ruled out this risk factor.
    \end{itemize}
\end{itemize}

\medskip
\textbf{\# OUTPUT FORMAT REQUIREMENTS}
\begin{enumerate}[leftmargin=1.5em, label=\arabic*., noitemsep]
    \item \textbf{Soliloquy First, Summary Later}: Output the full personal soliloquy first.
    \item \textbf{Clear Separation}: After the soliloquy, use ``---'' as a separator.
    \item \textbf{Final Conclusion}: After the separator, list only the \textbf{accepted} risk factors to form the final report.
\end{enumerate}

\medskip
\textbf{\# INPUT DATA}
\begin{itemize}[leftmargin=1.5em, noitemsep]
    \item \textbf{raw data}: \{Raw Data\}
    \item \textbf{accepted risks}: \{Accepted Risks Json\}
    \item \textbf{rejected risks}: \{Rejected Risks Json\}
\end{itemize}
\end{promptbox}

\section{A Case Study}
To provide an intuitive understanding of \textsc{Sherlock}'s operational mechanism, we present a representative execution trace in Table \ref{tab:appendix_lls_full_trace}. For the sake of data privacy and confidentiality, the displayed data has been de-identified, and we present a representative subset of the reasoning process rather than the full raw transaction logs.

\begin{table*}[htbp]
\centering
\caption{End-to-end reasoning trace of \textsc{Sherlock} in an LLS scenario.}
\label{tab:appendix_lls_full_trace}
\small
\renewcommand{\arraystretch}{1.8}
\begin{tabular}{p{3cm}|p{13cm}}
\toprule
\rowcolor[gray]{.95} \textbf{System Stage} & \textbf{Trace: Data Stream \& Reasoning (CoT)} \\
\midrule
\textbf{Raw Input} \newline (Serialization) & 
\texttt{\#\#\# Tabular Content} \newline 
\texttt{\{ "order\_id": "O\_99281", "vendor\_id": "V\_771", "pin": "U\_052", "biz\_line": "LLS", "item\_name": "Iced Americano XL", ... \}} \newline 
\texttt{\#\#\# Graph Context} \newline 
\texttt{(U\_052, U\_053, IP\_10.x), (U\_053, U\_054, IP\_10.x), (U\_054, U\_055, IP\_10.x), ..., (U\_059, U\_060, IP\_10.x)} \newline 
\texttt{\#\#\# Textual Context} \newline 
\texttt{"Address: 22F, Global Science \& Technology Tower B. Comment: 'Please deliver to the front desk of the R\&D department.'"} \\
\midrule
\textbf{Augmented Input} \newline (1st-Pass Retrieval) & 
\textbf{Retrieved Domain Term:} \newline
\texttt{\{ "term": "LLS", "definition": "Local Life Service; an innovative business model where online orders are fulfilled through offline in-store services. It is fundamentally characterized by the physical proximity of users and merchants." \}} \\
\midrule
\textbf{Preliminary Output} \newline (Initial Analysis) & 
\textbf{CoT:} \newline
Analyzing the case within the LLS context, the primary anomaly is the extreme network-level density where multiple accounts (U\_052--U\_060) share an identical IP gateway. In standard security paradigms, high-density IP clustering is a classic indicator of device farms or botnets used to simulate fake orders. Given the volume of accounts originating from a single network exit in a narrow time window, I suspect coordinated adversarial activity. Therefore, I initially flag this clustering as a malicious signal intended to bypass platform detection. \newline
\textbf{Output:} \newline
\texttt{\{ "judgment": "malicious", "risk\_factors": [\{"title": "IP Clustering", "desc": "High-density account aggregation on a single network exit point."\}] \}} \\
\midrule
\textbf{Retrieved Knowledge} \newline (2nd-Pass Retrieval) & 
\textbf{Retrieved Business Prior:} \newline
\texttt{\{ "risk\_factor": "IP clustering", "business\_logic": "In LLS scenarios, user groups are naturally concentrated near the merchant's physical location. In venues such as corporate towers or schools, users typically access the platform via shared Wi-Fi or internal NAT gateways, resulting in legitimate IP clustering. This pattern is a normal characteristic of localized high-density demand." \}} \\
\midrule
\textbf{Refined Output} \newline (Reflect \& Refine) & 
\textbf{CoT:} \newline
I am re-evaluating the preliminary malicious judgment by applying the "IP clustering" business prior specific to LLS environments. While the network density is objectively high, the textual context identifies the location as "Global Science \& Technology Tower B". This aligns with the "shared Wi-Fi" logic found in corporate settings, where large groups of employees in the same physical building place orders (such as coffee) through a unified internal gateway. The initial suspicion of a device farm is ruled out because the merchant-centered locality and the nature of the LLS service provide a high-fidelity benign explanation. This clustering is not a sign of collusion but a normal manifestation of concentrated demand from a single enterprise network. \newline
\textbf{Output:} \newline
\texttt{\{ "judgment": "benign", "risk\_factors": [] \}} \\
\bottomrule
\end{tabular}
\end{table*}

%% file: main.bbl

\begin{thebibliography}{44}


\ifx \showCODEN    \undefined \def \showCODEN     #1{\unskip}     \fi
\ifx \showISBNx    \undefined \def \showISBNx     #1{\unskip}     \fi
\ifx \showISBNxiii \undefined \def \showISBNxiii  #1{\unskip}     \fi
\ifx \showISSN     \undefined \def \showISSN      #1{\unskip}     \fi
\ifx \showLCCN     \undefined \def \showLCCN      #1{\unskip}     \fi
\ifx \shownote     \undefined \def \shownote      #1{#1}          \fi
\ifx \showarticletitle \undefined \def \showarticletitle #1{#1}   \fi
\ifx \showURL      \undefined \def \showURL       {\relax}        \fi
\providecommand\bibfield[2]{#2}
\providecommand\bibinfo[2]{#2}
\providecommand\natexlab[1]{#1}
\providecommand\showeprint[2][]{arXiv:#2}

\bibitem[Ali et~al\mbox{.}(2022)]%
        {ali2022financial}
\bibfield{author}{\bibinfo{person}{Abdulalem Ali}, \bibinfo{person}{Shukor Abd~Razak}, \bibinfo{person}{Siti~Hajar Othman}, \bibinfo{person}{Taiseer Abdalla~Elfadil Eisa}, \bibinfo{person}{Arafat Al-Dhaqm}, \bibinfo{person}{Maged Nasser}, \bibinfo{person}{Tusneem Elhassan}, \bibinfo{person}{Hashim Elshafie}, {and} \bibinfo{person}{Abdu Saif}.} \bibinfo{year}{2022}\natexlab{}.
\newblock \showarticletitle{Financial fraud detection based on machine learning: a systematic literature review}.
\newblock \bibinfo{journal}{\emph{Applied Sciences}} \bibinfo{volume}{12}, \bibinfo{number}{19} (\bibinfo{year}{2022}), \bibinfo{pages}{9637}.
\newblock


\bibitem[Cao et~al\mbox{.}(2019)]%
        {fraudant}
\bibfield{author}{\bibinfo{person}{Shaosheng Cao}, \bibinfo{person}{Xinxing Yang}, \bibinfo{person}{Cen Chen}, \bibinfo{person}{Jun Zhou}, \bibinfo{person}{Xiaolong Li}, {and} \bibinfo{person}{Yuan Qi}.} \bibinfo{year}{2019}\natexlab{}.
\newblock \showarticletitle{TitAnt: Online Real-time Transaction Fraud Detection in Ant Financial}.
\newblock \bibinfo{journal}{\emph{Proc. {VLDB} Endow.}} \bibinfo{volume}{12}, \bibinfo{number}{12} (\bibinfo{year}{2019}), \bibinfo{pages}{2082--2093}.
\newblock


\bibitem[Caron et~al\mbox{.}(2013)]%
        {caron2013comprehensive}
\bibfield{author}{\bibinfo{person}{Filip Caron}, \bibinfo{person}{Jan Vanthienen}, {and} \bibinfo{person}{Bart Baesens}.} \bibinfo{year}{2013}\natexlab{}.
\newblock \showarticletitle{Comprehensive rule-based compliance checking and risk management with process mining}.
\newblock \bibinfo{journal}{\emph{Decis. Support Syst.}} \bibinfo{volume}{54}, \bibinfo{number}{3} (\bibinfo{year}{2013}), \bibinfo{pages}{1357--1369}.
\newblock


\bibitem[Chen et~al\mbox{.}(2025)]%
        {chen2024bge}
\bibfield{author}{\bibinfo{person}{Jianlv Chen}, \bibinfo{person}{Shitao Xiao}, \bibinfo{person}{Peitian Zhang}, \bibinfo{person}{Kun Luo}, \bibinfo{person}{Defu Lian}, {and} \bibinfo{person}{Zheng Liu}.} \bibinfo{year}{2025}\natexlab{}.
\newblock \bibinfo{booktitle}{\emph{M3-Embedding: Multi-lingual, multi-functionality, multi-granularity text embeddings through self-knowledge distillation}}.
\newblock
\showeprint[arxiv]{2402.03216}~[cs.CL]


\bibitem[Chen et~al\mbox{.}(2024)]%
        {chen2024general2specialized}
\bibfield{author}{\bibinfo{person}{Kaidi Chen}, \bibinfo{person}{Ben Chen}, \bibinfo{person}{Dehong Gao}, \bibinfo{person}{Huangyu Dai}, \bibinfo{person}{Wen Jiang}, \bibinfo{person}{Wei Ning}, \bibinfo{person}{Shanqing Yu}, \bibinfo{person}{Libin Yang}, {and} \bibinfo{person}{Xiaoyan Cai}.} \bibinfo{year}{2024}\natexlab{}.
\newblock \showarticletitle{General2Specialized LLMs Translation for E-commerce}. In \bibinfo{booktitle}{\emph{{WWW} (Companion Volume)}}. \bibinfo{publisher}{{ACM}}, \bibinfo{pages}{670--673}.
\newblock


\bibitem[Cheng et~al\mbox{.}(2025)]%
        {logicreasoning}
\bibfield{author}{\bibinfo{person}{Fengxiang Cheng}, \bibinfo{person}{Haoxuan Li}, \bibinfo{person}{Fenrong Liu}, \bibinfo{person}{Robert van Rooij}, \bibinfo{person}{Kun Zhang}, {and} \bibinfo{person}{Zhouchen Lin}.} \bibinfo{year}{2025}\natexlab{}.
\newblock \showarticletitle{Empowering LLMs with Logical Reasoning: {A} Comprehensive Survey}. In \bibinfo{booktitle}{\emph{{IJCAI}}}. \bibinfo{publisher}{ijcai.org}, \bibinfo{pages}{10400--10408}.
\newblock


\bibitem[DeepSeek{-}AI(2025)]%
        {deepseek-r1}
\bibfield{author}{\bibinfo{person}{DeepSeek{-}AI}.} \bibinfo{year}{2025}\natexlab{}.
\newblock \bibinfo{booktitle}{\emph{DeepSeek-R1: Incentivizing Reasoning Capability in LLMs via Reinforcement Learning}}.
\newblock
\showeprint[arxiv]{2501.12948}~[cs.CL]


\bibitem[Esenogho et~al\mbox{.}(2022)]%
        {fraud2022deep2}
\bibfield{author}{\bibinfo{person}{Ebenezer Esenogho}, \bibinfo{person}{Domor~Mienye Ibomoiye}, \bibinfo{person}{Theo~G. Swart}, \bibinfo{person}{Kehinde~D. Aruleba}, {and} \bibinfo{person}{George Obaido}.} \bibinfo{year}{2022}\natexlab{}.
\newblock \showarticletitle{A Neural Network Ensemble With Feature Engineering for Improved Credit Card Fraud Detection}.
\newblock \bibinfo{journal}{\emph{{IEEE} Access}}  \bibinfo{volume}{10} (\bibinfo{year}{2022}), \bibinfo{pages}{16400--16407}.
\newblock


\bibitem[Feng(2025)]%
        {feng2025hybrid}
\bibfield{author}{\bibinfo{person}{Peter Feng}.} \bibinfo{year}{2025}\natexlab{}.
\newblock \showarticletitle{Hybrid BiLSTM-Transformer Model for Identifying Fraudulent Transactions in Financial Systems}.
\newblock \bibinfo{journal}{\emph{Journal of Computer Science and Software Applications}} \bibinfo{volume}{5}, \bibinfo{number}{3} (\bibinfo{year}{2025}).
\newblock


\bibitem[Gohel et~al\mbox{.}(2021)]%
        {gohel2021explainable}
\bibfield{author}{\bibinfo{person}{Prashant Gohel}, \bibinfo{person}{Priyanka Singh}, {and} \bibinfo{person}{Manoranjan Mohanty}.} \bibinfo{year}{2021}\natexlab{}.
\newblock \bibinfo{booktitle}{\emph{Explainable AI: current status and future directions}}.
\newblock
\showeprint[arxiv]{2107.07045}~[cs.LG]


\bibitem[Hamilton et~al\mbox{.}(2017)]%
        {SAGE}
\bibfield{author}{\bibinfo{person}{William~L. Hamilton}, \bibinfo{person}{Zhitao Ying}, {and} \bibinfo{person}{Jure Leskovec}.} \bibinfo{year}{2017}\natexlab{}.
\newblock \showarticletitle{Inductive Representation Learning on Large Graphs}. In \bibinfo{booktitle}{\emph{{NIPS}}}. \bibinfo{pages}{1024--1034}.
\newblock


\bibitem[Huang et~al\mbox{.}(2025)]%
        {huang2025can}
\bibfield{author}{\bibinfo{person}{Tairan Huang}, \bibinfo{person}{Yili Wang}, \bibinfo{person}{Qiutong Li}, \bibinfo{person}{Changlong He}, {and} \bibinfo{person}{Jianliang Gao}.} \bibinfo{year}{2025}\natexlab{}.
\newblock \showarticletitle{Can LLMs Find Fraudsters? Multi-level {LLM} Enhanced Graph Fraud Detection}. In \bibinfo{booktitle}{\emph{{ACM} Multimedia}}. \bibinfo{publisher}{{ACM}}, \bibinfo{pages}{1530--1538}.
\newblock


\bibitem[IBRAHIM et~al\mbox{.}(2025)]%
        {ibrahim2025rule}
\bibfield{author}{\bibinfo{person}{Bisallah~H IBRAHIM}, \bibinfo{person}{Habiba~U SALIHU}, {and} \bibinfo{person}{Yusuf~A ALESHINLOYE}.} \bibinfo{year}{2025}\natexlab{}.
\newblock \showarticletitle{Rule-Based Approach to e-Commerce Fraud Detection}.
\newblock \bibinfo{journal}{\emph{UNIABUJA Journal of Engineering and Technology (UJET)}} \bibinfo{volume}{2}, \bibinfo{number}{1} (\bibinfo{year}{2025}), \bibinfo{pages}{196--204}.
\newblock


\bibitem[Ji et~al\mbox{.}(2023)]%
        {ji2023towards}
\bibfield{author}{\bibinfo{person}{Ziwei Ji}, \bibinfo{person}{Tiezheng Yu}, \bibinfo{person}{Yan Xu}, \bibinfo{person}{Nayeon Lee}, \bibinfo{person}{Etsuko Ishii}, {and} \bibinfo{person}{Pascale Fung}.} \bibinfo{year}{2023}\natexlab{}.
\newblock \bibinfo{booktitle}{\emph{Towards mitigating hallucination in large language models via self-reflection}}.
\newblock
\showeprint[arxiv]{2310.06271}~[cs.CL]


\bibitem[Jiao et~al\mbox{.}(2024)]%
        {logicllm}
\bibfield{author}{\bibinfo{person}{Fangkai Jiao}, \bibinfo{person}{Zhiyang Teng}, \bibinfo{person}{Bosheng Ding}, \bibinfo{person}{Zhengyuan Liu}, \bibinfo{person}{Nancy~F. Chen}, {and} \bibinfo{person}{Shafiq Joty}.} \bibinfo{year}{2024}\natexlab{}.
\newblock \showarticletitle{Exploring Self-supervised Logic-enhanced Training for Large Language Models}. In \bibinfo{booktitle}{\emph{{NAACL-HLT}}}. \bibinfo{publisher}{Association for Computational Linguistics}, \bibinfo{pages}{926--941}.
\newblock


\bibitem[Lewis et~al\mbox{.}(2020a)]%
        {lewis2020retrieval}
\bibfield{author}{\bibinfo{person}{Patrick Lewis}, \bibinfo{person}{Ethan Perez}, \bibinfo{person}{Aleksandra Piktus}, \bibinfo{person}{Fabio Petroni}, \bibinfo{person}{Vladimir Karpukhin}, \bibinfo{person}{Naman Goyal}, \bibinfo{person}{Heinrich K{\"{u}}ttler}, \bibinfo{person}{Mike Lewis}, \bibinfo{person}{Wen{-}tau Yih}, \bibinfo{person}{Tim Rockt{\"{a}}schel}, \bibinfo{person}{Sebastian Riedel}, {and} \bibinfo{person}{Douwe Kiela}.} \bibinfo{year}{2020}\natexlab{a}.
\newblock \showarticletitle{Retrieval-Augmented Generation for Knowledge-Intensive {NLP} Tasks}. In \bibinfo{booktitle}{\emph{NeurIPS}}.
\newblock


\bibitem[Lewis et~al\mbox{.}(2020b)]%
        {RAG}
\bibfield{author}{\bibinfo{person}{Patrick Lewis}, \bibinfo{person}{Ethan Perez}, \bibinfo{person}{Aleksandra Piktus}, \bibinfo{person}{Fabio Petroni}, \bibinfo{person}{Vladimir Karpukhin}, \bibinfo{person}{Naman Goyal}, \bibinfo{person}{Heinrich K{\"{u}}ttler}, \bibinfo{person}{Mike Lewis}, \bibinfo{person}{Wen{-}tau Yih}, \bibinfo{person}{Tim Rockt{\"{a}}schel}, \bibinfo{person}{Sebastian Riedel}, {and} \bibinfo{person}{Douwe Kiela}.} \bibinfo{year}{2020}\natexlab{b}.
\newblock \showarticletitle{Retrieval-Augmented Generation for Knowledge-Intensive {NLP} Tasks}. In \bibinfo{booktitle}{\emph{NeurIPS}}.
\newblock


\bibitem[Li et~al\mbox{.}(2020)]%
        {li2019time}
\bibfield{author}{\bibinfo{person}{Longfei Li}, \bibinfo{person}{Ziqi Liu}, \bibinfo{person}{Chaochao Chen}, \bibinfo{person}{Ya-Lin Zhang}, \bibinfo{person}{Jun Zhou}, {and} \bibinfo{person}{Xiaolong Li}.} \bibinfo{year}{2020}\natexlab{}.
\newblock \bibinfo{booktitle}{\emph{A time attention based fraud transaction detection framework}}.
\newblock
\showeprint[arxiv]{1912.11760}~[cs.LG]


\bibitem[Li et~al\mbox{.}(2026)]%
        {li2025dgp}
\bibfield{author}{\bibinfo{person}{Yuan Li}, \bibinfo{person}{Jun Hu}, \bibinfo{person}{Bryan Hooi}, \bibinfo{person}{Bingsheng He}, {and} \bibinfo{person}{Cheng Chen}.} \bibinfo{year}{2026}\natexlab{}.
\newblock \showarticletitle{{DGP:} {A} Dual-Granularity Prompting Framework for Fraud Detection with Graph-Enhanced LLMs}. In \bibinfo{booktitle}{\emph{{AAAI}}}. \bibinfo{publisher}{{AAAI} Press}, \bibinfo{pages}{15171--15179}.
\newblock


\bibitem[Li et~al\mbox{.}(2024)]%
        {li2024ecomgpt}
\bibfield{author}{\bibinfo{person}{Yangning Li}, \bibinfo{person}{Shirong Ma}, \bibinfo{person}{Xiaobin Wang}, \bibinfo{person}{Shen Huang}, \bibinfo{person}{Chengyue Jiang}, \bibinfo{person}{Haitao Zheng}, \bibinfo{person}{Pengjun Xie}, \bibinfo{person}{Fei Huang}, {and} \bibinfo{person}{Yong Jiang}.} \bibinfo{year}{2024}\natexlab{}.
\newblock \showarticletitle{EcomGPT: Instruction-Tuning Large Language Models with Chain-of-Task Tasks for E-commerce}. In \bibinfo{booktitle}{\emph{{AAAI}}}. \bibinfo{publisher}{{AAAI} Press}, \bibinfo{pages}{18582--18590}.
\newblock


\bibitem[Liu et~al\mbox{.}(2025)]%
        {liu2025newworldwordembeddings}
\bibfield{author}{\bibinfo{person}{Zhu Liu}, \bibinfo{person}{Ying Liu}, \bibinfo{person}{KangYang Luo}, \bibinfo{person}{Cunliang Kong}, {and} \bibinfo{person}{Maosong Sun}.} \bibinfo{year}{2025}\natexlab{}.
\newblock \bibinfo{booktitle}{\emph{From the New World of Word Embeddings: A Comparative Study of Small-World Lexico-Semantic Networks in LLMs}}.
\newblock
\showeprint[arxiv]{2502.11380}~[cs.CL]


\bibitem[Luo et~al\mbox{.}(2024)]%
        {luo2024arena}
\bibfield{author}{\bibinfo{person}{Haipeng Luo}, \bibinfo{person}{Qingfeng Sun}, \bibinfo{person}{Can Xu}, \bibinfo{person}{Pu Zhao}, \bibinfo{person}{Qingwei Lin}, \bibinfo{person}{Jianguang Lou}, \bibinfo{person}{Shifeng Chen}, \bibinfo{person}{Yansong Tang}, {and} \bibinfo{person}{Weizhu Chen}.} \bibinfo{year}{2024}\natexlab{}.
\newblock \bibinfo{booktitle}{\emph{Arena learning: Build data flywheel for llms post-training via simulated chatbot arena}}.
\newblock
\showeprint[arxiv]{2407.10627}~[cs.CL]


\bibitem[Luo et~al\mbox{.}(2025a)]%
        {luo2025chatrule}
\bibfield{author}{\bibinfo{person}{Linhao Luo}, \bibinfo{person}{Jiaxin Ju}, \bibinfo{person}{Bo Xiong}, \bibinfo{person}{Yuan{-}Fang Li}, \bibinfo{person}{Gholamreza Haffari}, {and} \bibinfo{person}{Shirui Pan}.} \bibinfo{year}{2025}\natexlab{a}.
\newblock \showarticletitle{ChatRule: Mining Logical Rules with Large Language Models for Knowledge Graph Reasoning}. In \bibinfo{booktitle}{\emph{{PAKDD} {(2)}}} \emph{(\bibinfo{series}{Lecture Notes in Computer Science})}. \bibinfo{publisher}{Springer}, \bibinfo{pages}{314--325}.
\newblock


\bibitem[Luo et~al\mbox{.}(2025b)]%
        {luo2025fraud}
\bibfield{author}{\bibinfo{person}{RuiHan Luo}, \bibinfo{person}{Nanxi Wang}, {and} \bibinfo{person}{Xiaotong Zhu}.} \bibinfo{year}{2025}\natexlab{b}.
\newblock \bibinfo{booktitle}{\emph{Fraud detection and risk assessment of online payment transactions on e-commerce platforms based on LLM and GCN frameworks}}.
\newblock
\showeprint[arxiv]{2509.09928}~[cs.CE]


\bibitem[Mecklenburg et~al\mbox{.}(2024)]%
        {knowledge-inject}
\bibfield{author}{\bibinfo{person}{Nick Mecklenburg}, \bibinfo{person}{Yiyou Lin}, \bibinfo{person}{Xiaoxiao Li}, \bibinfo{person}{Daniel Holstein}, \bibinfo{person}{Leonardo~O. Nunes}, \bibinfo{person}{Sara Malvar}, \bibinfo{person}{Bruno Silva}, \bibinfo{person}{Ranveer Chandra}, \bibinfo{person}{Vijay Aski}, \bibinfo{person}{Pavan Kumar~Reddy Yannam}, \bibinfo{person}{Tolga Aktas}, {and} \bibinfo{person}{Todd Hendry}.} \bibinfo{year}{2024}\natexlab{}.
\newblock \bibinfo{booktitle}{\emph{Injecting New Knowledge into Large Language Models via Supervised Fine-Tuning}}.
\newblock
\showeprint[arxiv]{2404.00213}~[cs.CL]


\bibitem[Motie and Raahemi(2024)]%
        {fraud-recent-3}
\bibfield{author}{\bibinfo{person}{Soroor Motie} {and} \bibinfo{person}{Bijan Raahemi}.} \bibinfo{year}{2024}\natexlab{}.
\newblock \showarticletitle{Financial fraud detection using graph neural networks: {A} systematic review}.
\newblock \bibinfo{journal}{\emph{Expert Syst. Appl.}}  \bibinfo{volume}{240} (\bibinfo{year}{2024}), \bibinfo{pages}{122156}.
\newblock


\bibitem[Ouyang et~al\mbox{.}(2022)]%
        {ouyang2022training}
\bibfield{author}{\bibinfo{person}{Long Ouyang}, \bibinfo{person}{Jeffrey Wu}, \bibinfo{person}{Xu Jiang}, \bibinfo{person}{Diogo Almeida}, \bibinfo{person}{Carroll~L. Wainwright}, \bibinfo{person}{Pamela Mishkin}, \bibinfo{person}{Chong Zhang}, \bibinfo{person}{Sandhini Agarwal}, \bibinfo{person}{Katarina Slama}, \bibinfo{person}{Alex Ray}, \bibinfo{person}{John Schulman}, \bibinfo{person}{Jacob Hilton}, \bibinfo{person}{Fraser Kelton}, \bibinfo{person}{Luke Miller}, \bibinfo{person}{Maddie Simens}, \bibinfo{person}{Amanda Askell}, \bibinfo{person}{Peter Welinder}, \bibinfo{person}{Paul~F. Christiano}, \bibinfo{person}{Jan Leike}, {and} \bibinfo{person}{Ryan Lowe}.} \bibinfo{year}{2022}\natexlab{}.
\newblock \showarticletitle{Training language models to follow instructions with human feedback}. In \bibinfo{booktitle}{\emph{NeurIPS}}.
\newblock


\bibitem[Palen{-}Michel et~al\mbox{.}(2024)]%
        {e-com-llm-1}
\bibfield{author}{\bibinfo{person}{Chester Palen{-}Michel}, \bibinfo{person}{Ruixiang Wang}, \bibinfo{person}{Yipeng Zhang}, \bibinfo{person}{David Yu}, \bibinfo{person}{Canran Xu}, {and} \bibinfo{person}{Zhe Wu}.} \bibinfo{year}{2024}\natexlab{}.
\newblock \bibinfo{booktitle}{\emph{Investigating {LLM} Applications in E-Commerce}}.
\newblock
\showeprint[arxiv]{2408.12779}~[cs.CL]


\bibitem[Peng et~al\mbox{.}(2024)]%
        {peng2024ecellm}
\bibfield{author}{\bibinfo{person}{Bo Peng}, \bibinfo{person}{Xinyi Ling}, \bibinfo{person}{Ziru Chen}, \bibinfo{person}{Huan Sun}, {and} \bibinfo{person}{Xia Ning}.} \bibinfo{year}{2024}\natexlab{}.
\newblock \showarticletitle{eCeLLM: Generalizing Large Language Models for E-commerce from Large-scale, High-quality Instruction Data}. In \bibinfo{booktitle}{\emph{{ICML}}} \emph{(\bibinfo{series}{Proceedings of Machine Learning Research})}. \bibinfo{publisher}{{PMLR} / OpenReview.net}, \bibinfo{pages}{40215--40257}.
\newblock


\bibitem[Rafailov et~al\mbox{.}(2023)]%
        {rafailov2023direct}
\bibfield{author}{\bibinfo{person}{Rafael Rafailov}, \bibinfo{person}{Archit Sharma}, \bibinfo{person}{Eric Mitchell}, \bibinfo{person}{Christopher~D. Manning}, \bibinfo{person}{Stefano Ermon}, {and} \bibinfo{person}{Chelsea Finn}.} \bibinfo{year}{2023}\natexlab{}.
\newblock \showarticletitle{Direct Preference Optimization: Your Language Model is Secretly a Reward Model}. In \bibinfo{booktitle}{\emph{NeurIPS}}.
\newblock


\bibitem[Rao et~al\mbox{.}(2021)]%
        {rao2020xfraud}
\bibfield{author}{\bibinfo{person}{Susie~Xi Rao}, \bibinfo{person}{Shuai Zhang}, \bibinfo{person}{Zhichao Han}, \bibinfo{person}{Zitao Zhang}, \bibinfo{person}{Wei Min}, \bibinfo{person}{Zhiyao Chen}, \bibinfo{person}{Yinan Shan}, \bibinfo{person}{Yang Zhao}, {and} \bibinfo{person}{Ce Zhang}.} \bibinfo{year}{2021}\natexlab{}.
\newblock \showarticletitle{xFraud: Explainable Fraud Transaction Detection}.
\newblock \bibinfo{journal}{\emph{Proc. {VLDB} Endow.}} \bibinfo{volume}{15}, \bibinfo{number}{3} (\bibinfo{year}{2021}), \bibinfo{pages}{427--436}.
\newblock


\bibitem[Song et~al\mbox{.}(2019)]%
        {song2019cross}
\bibfield{author}{\bibinfo{person}{Bo Song}, \bibinfo{person}{Wei Yan}, {and} \bibinfo{person}{Tianjiao Zhang}.} \bibinfo{year}{2019}\natexlab{}.
\newblock \showarticletitle{Cross-border e-commerce commodity risk assessment using text mining and fuzzy rule-based reasoning}.
\newblock \bibinfo{journal}{\emph{Adv. Eng. Informatics}}  \bibinfo{volume}{40} (\bibinfo{year}{2019}), \bibinfo{pages}{69--80}.
\newblock


\bibitem[Team(2024)]%
        {dubey2024llama}
\bibfield{author}{\bibinfo{person}{Llama Team}.} \bibinfo{year}{2024}\natexlab{}.
\newblock \bibinfo{booktitle}{\emph{The Llama 3 Herd of Models}}.
\newblock
\showeprint[arxiv]{2407.21783}~[cs.AI]


\bibitem[Team(2025)]%
        {yang2025qwen3}
\bibfield{author}{\bibinfo{person}{Qwen Team}.} \bibinfo{year}{2025}\natexlab{}.
\newblock \bibinfo{booktitle}{\emph{Qwen3 Technical Report}}.
\newblock
\showeprint[arxiv]{2505.09388}~[cs.CL]


\bibitem[Velickovic et~al\mbox{.}(2018)]%
        {GAT}
\bibfield{author}{\bibinfo{person}{Petar Velickovic}, \bibinfo{person}{Guillem Cucurull}, \bibinfo{person}{Arantxa Casanova}, \bibinfo{person}{Adriana Romero}, \bibinfo{person}{Pietro Li{\`{o}}}, {and} \bibinfo{person}{Yoshua Bengio}.} \bibinfo{year}{2018}\natexlab{}.
\newblock \showarticletitle{Graph Attention Networks}. In \bibinfo{booktitle}{\emph{{ICLR} (Poster)}}. \bibinfo{publisher}{OpenReview.net}.
\newblock


\bibitem[Wang et~al\mbox{.}(2024)]%
        {wang2024taste}
\bibfield{author}{\bibinfo{person}{Yutong Wang}, \bibinfo{person}{Jiali Zeng}, \bibinfo{person}{Xuebo Liu}, \bibinfo{person}{Fandong Meng}, \bibinfo{person}{Jie Zhou}, {and} \bibinfo{person}{Min Zhang}.} \bibinfo{year}{2024}\natexlab{}.
\newblock \showarticletitle{TasTe: Teaching Large Language Models to Translate through Self-Reflection}. In \bibinfo{booktitle}{\emph{{ACL} {(1)}}}. \bibinfo{publisher}{Association for Computational Linguistics}, \bibinfo{pages}{6144--6158}.
\newblock


\bibitem[Wang et~al\mbox{.}(2025)]%
        {wang2024bootstrapping}
\bibfield{author}{\bibinfo{person}{Zun Wang}, \bibinfo{person}{Jialu Li}, \bibinfo{person}{Yicong Hong}, \bibinfo{person}{Songze Li}, \bibinfo{person}{Kunchang Li}, \bibinfo{person}{Shoubin Yu}, \bibinfo{person}{Yi Wang}, \bibinfo{person}{Yu Qiao}, \bibinfo{person}{Yali Wang}, \bibinfo{person}{Mohit Bansal}, {and} \bibinfo{person}{Limin Wang}.} \bibinfo{year}{2025}\natexlab{}.
\newblock \showarticletitle{Bootstrapping Language-Guided Navigation Learning with Self-Refining Data Flywheel}. In \bibinfo{booktitle}{\emph{{ICLR}}}. \bibinfo{publisher}{OpenReview.net}.
\newblock


\bibitem[Xu et~al\mbox{.}(2020)]%
        {e-com-KG}
\bibfield{author}{\bibinfo{person}{Da Xu}, \bibinfo{person}{Chuanwei Ruan}, \bibinfo{person}{Evren K{\"{o}}rpeoglu}, \bibinfo{person}{Sushant Kumar}, {and} \bibinfo{person}{Kannan Achan}.} \bibinfo{year}{2020}\natexlab{}.
\newblock \showarticletitle{Product Knowledge Graph Embedding for E-commerce}. In \bibinfo{booktitle}{\emph{{WSDM}}}. \bibinfo{publisher}{{ACM}}, \bibinfo{pages}{672--680}.
\newblock


\bibitem[Yang et~al\mbox{.}(2025a)]%
        {yang2025flag}
\bibfield{author}{\bibinfo{person}{Chengdong Yang}, \bibinfo{person}{Hongrui Liu}, \bibinfo{person}{Daixin Wang}, \bibinfo{person}{Zhiqiang Zhang}, \bibinfo{person}{Cheng Yang}, {and} \bibinfo{person}{Chuan Shi}.} \bibinfo{year}{2025}\natexlab{a}.
\newblock \showarticletitle{{FLAG:} Fraud Detection with LLM-enhanced Graph Neural Network}. In \bibinfo{booktitle}{\emph{{KDD} {(2)}}}. \bibinfo{publisher}{{ACM}}, \bibinfo{pages}{5150--5160}.
\newblock


\bibitem[Yang et~al\mbox{.}(2025b)]%
        {yang2025fraudr1}
\bibfield{author}{\bibinfo{person}{Shu Yang}, \bibinfo{person}{Shenzhe Zhu}, \bibinfo{person}{Zeyu Wu}, \bibinfo{person}{Keyu Wang}, \bibinfo{person}{Junchi Yao}, \bibinfo{person}{Junchao Wu}, \bibinfo{person}{Lijie Hu}, \bibinfo{person}{Mengdi Li}, \bibinfo{person}{Derek~F. Wong}, {and} \bibinfo{person}{Di Wang}.} \bibinfo{year}{2025}\natexlab{b}.
\newblock \showarticletitle{Fraud-R1 : {A} Multi-Round Benchmark for Assessing the Robustness of {LLM} Against Augmented Fraud and Phishing Inducements}. In \bibinfo{booktitle}{\emph{{ACL} (Findings)}} \emph{(\bibinfo{series}{Findings of {ACL}})}. \bibinfo{publisher}{Association for Computational Linguistics}, \bibinfo{pages}{4374--4420}.
\newblock


\bibitem[Yao et~al\mbox{.}(2025)]%
        {yao2025your}
\bibfield{author}{\bibinfo{person}{Junchi Yao}, \bibinfo{person}{Jianhua Xu}, \bibinfo{person}{Tianyu Xin}, \bibinfo{person}{Ziyi Wang}, \bibinfo{person}{Shenzhe Zhu}, \bibinfo{person}{Shu Yang}, {and} \bibinfo{person}{Di Wang}.} \bibinfo{year}{2025}\natexlab{}.
\newblock \bibinfo{booktitle}{\emph{Is your llm-based multi-agent a reliable real-world planner? exploring fraud detection in travel planning}}.
\newblock
\showeprint[arxiv]{2505.16557}~[cs.MA]


\bibitem[Yin et~al\mbox{.}(2022)]%
        {yin2022behavioral}
\bibfield{author}{\bibinfo{person}{Hang Yin}, \bibinfo{person}{Zitao Zhang}, \bibinfo{person}{Zhurong Wang}, \bibinfo{person}{Yilmazcan {\"{O}}zyurt}, \bibinfo{person}{Weiming Liang}, \bibinfo{person}{Wenyu Dong}, \bibinfo{person}{Yang Zhao}, {and} \bibinfo{person}{Yinan Shan}.} \bibinfo{year}{2022}\natexlab{}.
\newblock \showarticletitle{Behavioral graph fraud detection in E-commerce}. In \bibinfo{booktitle}{\emph{{ICDM} (Workshops)}}. \bibinfo{publisher}{{IEEE}}, \bibinfo{pages}{1--8}.
\newblock


\bibitem[Yu et~al\mbox{.}(2023)]%
        {fraud-recent-1}
\bibfield{author}{\bibinfo{person}{Jianke Yu}, \bibinfo{person}{Hanchen Wang}, \bibinfo{person}{Xiaoyang Wang}, \bibinfo{person}{Zhao Li}, \bibinfo{person}{Lu Qin}, \bibinfo{person}{Wenjie Zhang}, \bibinfo{person}{Jian Liao}, {and} \bibinfo{person}{Ying Zhang}.} \bibinfo{year}{2023}\natexlab{}.
\newblock \showarticletitle{Group-based Fraud Detection Network on e-Commerce Platforms}. In \bibinfo{booktitle}{\emph{{KDD}}}. \bibinfo{publisher}{{ACM}}, \bibinfo{pages}{5463--5475}.
\newblock


\bibitem[Zhou et~al\mbox{.}(2023)]%
        {lima}
\bibfield{author}{\bibinfo{person}{Chunting Zhou}, \bibinfo{person}{Pengfei Liu}, \bibinfo{person}{Puxin Xu}, \bibinfo{person}{Srinivasan Iyer}, \bibinfo{person}{Jiao Sun}, \bibinfo{person}{Yuning Mao}, \bibinfo{person}{Xuezhe Ma}, \bibinfo{person}{Avia Efrat}, \bibinfo{person}{Ping Yu}, \bibinfo{person}{Lili Yu}, \bibinfo{person}{Susan Zhang}, \bibinfo{person}{Gargi Ghosh}, \bibinfo{person}{Mike Lewis}, \bibinfo{person}{Luke Zettlemoyer}, {and} \bibinfo{person}{Omer Levy}.} \bibinfo{year}{2023}\natexlab{}.
\newblock \showarticletitle{{LIMA:} Less Is More for Alignment}. In \bibinfo{booktitle}{\emph{NeurIPS}}.
\newblock


\end{thebibliography}
